\documentclass[sigconf]{acmart}
\AtBeginDocument{%
  }
\usepackage{hyperref}
\copyrightyear{2025}
\acmYear{2025}
\setcopyright{rightsretained}
\acmConference[CHI EA '25]{Extended Abstracts of the CHI Conference on Human Factors in Computing Systems}{April 26-May 1, 2025}{Yokohama, Japan}
\acmBooktitle{Extended Abstracts of the CHI Conference on Human Factors in Computing Systems (CHI EA '25), April 26-May 1, 2025, Yokohama, Japan}\acmDOI{10.1145/3706599.3720032}
\acmISBN{979-8-4007-1395-8/2025/04}

\begin{document}

\title{``We need to avail ourselves of [GenAI] to enhance knowledge distribution'': Empowering Older Adults through GenAI Literacy}

\author{Eunhye Grace Ko}
\email{eko@utexas.edu}
\orcid{1234-5678-9012}
\affiliation{%
\department{School of Information}
  \institution{University of Texas at Austin}
  \city{Austin}
  \state{Texas}
  \country{USA}
}

\author{Shaini Nanayakkara}
\email{shaini.nanayakkara@utexas.edu}
\orcid{1234-5678-9012}
\affiliation{%
\department{School of Information}
  \institution{University of Texas at Austin}
  \city{Austin}
  \state{Texas}
  \country{USA}
}

\author{Earl W. Huff, Jr.}
\email{ewhuff@utexas.edu}
\orcid{1234-5678-9012}
\affiliation{%
\department{School of Information}
  \institution{University of Texas at Austin}
  \city{Austin}
  \state{Texas}
  \country{USA}
}

\renewcommand{\shortauthors}{Ko et al.}

\begin{abstract}
  As generative AI (GenAI) becomes increasingly ubiquitous, it is crucial to equip users, particularly vulnerable populations like older adults (65+), with the knowledge to understand its benefits and potential risks. Older adults often face greater reservations about adopting emerging technologies and require tailored literacy support. Using a mixed methods approach, this study examines strategies for delivering GenAI literacy to older adults through a chatbot named Litti, evaluating its impact on their Al literacy (knowledge, safety, and ethical use). The quantitative data showed a trend toward improved AI literacy, though the results were not statistically significant. However, the qualitative interviews revealed diverse levels of familiarity with generative AI, along with a strong desire to learn more. Qualitative findings also show that although Litti provided a positive learning experience, it did not significantly enhance participants’ trust or sense of safety regarding GenAI. This exploratory case study highlights the challenges and opportunities in designing AI literacy education for the rapidly growing older adult population.
\end{abstract}




\maketitle

\section{Introduction}
The increasing prevalence of Generative AI (GenAI) in various aspects of modern society has raised important questions about older adults' ability to understand and engage with these technologies. As a rapidly growing demographic \cite{crampton_population_2011}, older adults stand to benefit significantly from GenAI technologies that are transforming healthcare \cite{santos_multi-agent_2025} and introducing conversational systems that provide personalized support to combat social isolation and offer emotional companionship \cite{alessa_towards_2023}, thereby enhancing the agency of older adults. However, to fully harness these benefits and mitigate potential risks such as scams, it is vital for older adults to develop sufficient AI literacy to engage with these tools effectively and safely. Furthermore, it is important for older adults to be aware of societal and ethical implications these new technologies will bring to their near future.

To fully harness these benefits and mitigate potential risks, it is crucial for older adults to develop AI literacy. Thus, this study aims to address the existing gaps in AI literacy among older adults by conducting an empirical investigation using an AI chatbot, ‘Litti,’ specifically designed to deliver AI literacy education for older adults. By providing hands-on experience with Generative AI, our intervention focuses on enhancing their understanding of AI capabilities, addressing AI safety and ethical concerns, and fostering trust in these technologies. The findings will benefit not only the scientific community, particularly AI developers and researchers working on the intersection of AI accessibility and ethics, but also broader knowledge-sharing communities. By sharing insights into practical strategies for developing GenAI literacy for older adults, this study contributes to the development of inclusive AI systems and a more equitable technological landscape for all demographics. Thereby, we ask: "How does an AI literacy education chatbot impact older adults' AI literacy, including their knowledge, safe usage, ethical understanding, and trust in Generative AI??''
\section{Related Work}
\subsection{AI Literacy Education}
AI literacy, as defined by Long and Magerko, encompasses the ability to critically evaluate AI technologies, effectively interact with AI systems, and leverage them as tools in both domestic and professional settings \cite{long_what_2020}. AI literacy has become a critical competency in modern society, enabling individuals to understand and navigate the complexities of AI technologies. AI literacy education is progressively being integrated across different educational levels and disciplines. AI literacy education spans from K-12 programs to higher education and corporate upskilling initiatives, equipping diverse populations—including older adults—with foundational knowledge to improve human-AI communication and advance human-in-the-loop AI systems \cite{chetty_ai_2023, gomstyn_ai_2025, laupichler_artificial_2022, long_family_2022}  

Despite the growing integration of AI literacy into various educational contexts, a significant gap remains in addressing the needs of vulnerable populations such as older adults. The willingness of older adults to embrace new technologies is shaped by three key factors: perceived usefulness, technical literacy, and personal apprehensions about engaging with digital systems \cite{shandilya_understanding_2024} . These underscore the critical importance of AI literacy education for this demographic. However, research specifically exploring AI literacy for older adults is limited and preliminary, leaving a crucial gap in understanding how to effectively empower this population to navigate and benefit from AI technologies.

\subsection{AI Literacy and Education for Older Adults}
By 2050, the global population aged 65 and older will double to 1.5 billion, with 1 in 6 people aged 65 or over, and older adults are the fastest-growing U.S. demographic \cite{ortman_aging_2014, united_nations_world_2020}. While this increase in life expectancy is celebrated, older adults remain the largest demographic excluded from the digital realm, with many unable to access or benefit from online services due to insufficient digital literacy \cite{lu_digital_2022}. Older adults face unique challenges in adopting AI tools, including fears of scams, fraudulent activities, or malicious attacks, and with AI advancements increasing the frequency, potency, and effectiveness of scams \cite{mirsky_threat_2023} . In addition, age-related cognitive and visual impairments can reduce decision making abilities, making older adults more vulnerable to these risks [48]. Consequently, awareness and education about AI are crucial to mitigating these vulnerabilities \cite{zhang_artificial_2019} . By fostering AI literacy, older adults can develop the trust and knowledge needed to navigate AI-enhanced systems safely and effectively \cite{dhimolea_supporting_2022}.

Current research on AI literacy among older adults primarily focuses on evaluating their existing knowledge, experiences, and perceptions of AI technologies through surveys, interviews, and systematic reviews, often using examples like chatbots, Google Maps, and Alexa to contextualize and assess self-reported knowledge and usage patterns \cite{kaur_exploring_2023, shandilya_understanding_2024}. While these studies provide valuable insights, they lack educational intervention directly aimed at enhancing AI literacy \cite{oh_measurement_2021}.

The current literature gap lies in the absence of empirical research designed to provide older adults with practical skills to engage effectively with AI technologies, specifically focusing on GenAI. While general AI literacy and its implications have been explored across various demographics, little attention has been directed toward understanding how older adults perceive, interact with, and benefit from GenAI. This leaves a critical void in identifying effective strategies to support this demographic in navigating and leveraging GenAI effectively.

Our study seeks to address this gap by offering older adults hands-on experience with our chatbot, ‘Litti,’ a GenAI tool specifically designed to introduce them to the practical applications and capabilities of Generative AI which will not only allow participants to explore and interact with GenAI technologies but also help them build the skills necessary to engage with these tools. Through this, we aim to empower older adults to integrate GenAI into their daily lives, enhancing autonomy, confidence, and quality of life. Furthermore, we aim to lay groundwork for future studies on GenAI literacy for older adults.
\section{Intervention: Litti, the AI Literacy Education Chatbot}
We developed an AI literacy education chatbot, Litti, for older adults aged 65 and up. As there was a lack of a framework specifically developed for AI literacy education of older adults, we adapted the theoretical grounding provided by the Meta AI Literacy Scale (MAILS) framework developed \cite{carolus_mails_2023}. The MAILS framework was selected due to its comprehensive approach, which meaningfully extends previous work on the conceptualization and measurement of AI literacy. Key features of the MAILS include a strong theoretical foundation in the existing literature, a modular structure for flexible application, and rigorous psychometric testing evidence.To adapt the MAILS framework for our older adult population, we prioritized the sub-scales of ``Know \& AI,'' ``Detect AI,'' and ``AI Ethics'' over ``Create AI'' and ``AI Problem Solving.'' Tasks were designed around these selected sub-scales to ensure a well-rounded development of AI literacy in areas most relevant to their lives. This tailored approach was intended to effectively address the specific needs and capabilities of senior citizens, making the education both practical and impactful.

The chatbot was built using Claude, a large language model (LLM) from Anthropic \footnote{https://www.anthropic.com/claude}, and deployed through the FlowXO web-hosting platform \footnote{https://flowxo.com/}. The live prototype can be accessed \href{https://fxo.io/m/small-bike-6959}{HERE}.  The core of the chatbot’s design lies in the carefully crafted prompt that defines Litti’s persona and the guidelines for the conversational structure. Litti is positioned as an empathetic and engaging AI literacy instructor. After a brief introduction, Litti provides users with four tasks to complete. The prompt can be viewed \href{https://tinyurl.com/litti2025}{HERE}.

In the first task, “Exploring Generative AI for General Search,” participants are invited to use the chatbot to find quick information, such as answers to health-related questions, and learn about the benefits and limitations of this approach. The second task, “Exploring Generative AI for Personal Assistance,” allows participants to experience how the chatbot can provide personalized help with daily tasks, such as writing personalized birthday cards or creating individualized recipes with personal dietary preferences. The third task, “Emotional Companion for Mental Health Support,” explores the use of generative AI as an emotional companion, where participants can share their thoughts and concerns, and the chatbot responds with empathetic and supportive comments. In the fourth Task, “AI Safety,” participants learn about the safety concerns surrounding generative AI, such as voice cloning and deepfakes, and receive practical tips to identify and detect malicious use. The final task, “AI Ethics,” introduces participants to the ethical considerations around the societal impact of generative AI, including issues of fairness, inclusivity, and transparency. Participants are presented with real-world examples to illustrate the importance of critically evaluating the ethical implications of AI systems.
\section{Methods}
\subsection{Participants}
We recruited 12 older adults aged 75 and above (mean age = 81.1 years), with an equal distribution of men and women (6 each). The majority identified as White (n=10), with two participants each identifying as Hispanic, Latino, or Spanish origin, and Asian. Most participants were retired (n=10), while one was volunteering and another working part-time. For education, six held Master's degrees, three had Doctorates, two had Bachelor's degrees, and one had some college or an associate degree. Nine participants reported no impairments, while three reported auditory disabilities.

The older adults who participated in this study were familiar with using laptops and integrating technology into their daily lives. Since Litti was a web-based chatbot, participants interacted with it primarily via laptop devices, engaging in conversational interactions. Their comfort and proficiency levels with technology ranged from moderate to high, with some participants having experience using AI-driven tools such as Siri or Alexa in their daily lives. This prior experience likely facilitated their ability to quickly adapt to and effectively interact with the chatbot during the study.

Familiarity with AI varied among the participants; six rated their familiarity as 3 on a 5-point scale, two rated it as 4, one rated it as 1, and one as 5. In terms of AI usage frequency, five participants scored 5, two scored 4, one scored 2, and two scored 1. Technological self-efficacy also varied: three participants had average self-efficacy scores below 3.00, four had scores between 3.00 and 4.00, and three had scores above 4.00. The details of the participants are shown in the Appendix (Table \ref{tab:demo}). 

\subsection{Study Setup and Procedure}
The study took place on a single day given the availability of the space and participants at the study site. The study site was a senior assisted living center in an urban city in the Southwestern United States. We conducted the study in a room with four long tables, each with three chairs. We had four laptops (three Macbooks and one PC) and one Owl 360-degree camera. All procedures were conducted in an identical manner, and all sessions lasted approximately one hour. The research team introduced the study purpose, the participants’ involvement, and procedures.  As the institution’s IRB approved the study as exempt, only verbal consent was required from the participants. The three Macbooks were used to conduct the survey and chatbot interaction, with one computer at each of the three tables positioned so that participants could not see each other’s responses. We directed the participants to the computers and instructed them to take the pre-intervention survey. Once they completed the pre-survey, we directed them to the intervention through a web browser. The participants interacted with the chatbot, navigating the six steps presented one at a time. The interaction involved six tasks: 1) a brief explanation of generative AI, 2) a general search task, 3) exploring using AI as a personal assistant, 4) a health information-seeking task, 5) explore AI as an emotional support companion, and 6) learning about AI safety and ethics. After the intervention, participants completed a post-survey, structured like the pre-survey, to allow us to measure changes in their perceptions before and after working with the chatbot.   After the surveys and the chatbot, we reconvene with the participants for a post-study focus group to learn more about their experience with the chatbot and how it affected their perceptions of GenAI technology. We used a fourth computer to carry out the audio and video recording of the focus group, connecting with the OWL camera and recording through the Zoom conferencing software. Because of the OWL’s ability to isolate its focus to only those speaking, it allowed us to separate the conversations from the speakers during the transcription process. Each focus group lasted between 15-20 minutes. Once the focus group concluded, the study was completed.

\subsection{Survey \& Interview Questions}
The pre- and post-questionnaires were designed to evaluate participants’ AI literacy improvement after interacting with Litti. The questionnaire were developed adopting Meta AI Literacy Scale (MAILS) framework developed by Carolus et al. \cite{carolus_mails_2023}. The survey also included questions on their previous experiences with GenAI, technology self-efficacy, and general demographic questions. 

The focus group asked participants about their understanding of GenAI technology, what tools they were most eager to use in the future, feelings toward GenAI having used our intervention, ways that GenAI could be developed to support further older adults, and areas of improvement with the chatbot.
\section{Findings}
\subsection{Qualitative}
We transcribed all focus group recordings and analyzed them using Atlas.ti \footnote{https://atlasti.com/}, a qualitative analysis software. We used inductive coding on the transcripts, followed by thematic analysis.

\subsubsection{How Older Adults View AI vs. Generative AI}
Older adults perceive AI as a practical, reliable tool for daily tasks and safety, whereas GenAI is seen as more human-like and intellectually powerful. For many, AI serves as an assistive technology that enhances independence. P12 noted, “If I fell in the middle of my living room… I could say to Alexa, ‘call [the care facility],’ and it would make the call,” while also describing AI as an efficient personal assistant, stating, “It’s more like having your own personal butler or concierge… who will do things instantaneously.”
In contrast, GenAI is recognized for its ability to generate natural, human-like responses, as noted by P1, who stated, “It [GenAI] had a nice flow to it… it was not a mechanical response. It really sounded like a human had written it.” GenAI is also viewed as a knowledge-driven technology with transformative potential, particularly in education. P12 noted, “You could create an LLM of every public university in Texas… imagine what you could do in terms of educating students.”

\subsubsection{Prior Knowledge, Experience, and Need for AI literacy}
Participants' prior knowledge and experience with AI varied widely, from regular use of AI tools to limited or secondhand experiences of AI related incidents. P12 indicated a strong background in using AI, stating, "when I started, I taught courses here [senior facility] on using it [AI]" which contrasts with P10, who described themselves as a “total newbie” and expressed a desire to “learn more about it and do it better.” Others shared varying levels of familiarity with specific AI tools. P3 reported regular interaction with advanced AI systems, remarking, “I use Claude and Perplexity and ChatGPT. It’s all stuff I’m already familiar using pretty often.” In contrast, P6 shared indirect exposure to AI, stating, “I have a friend who uses it often... She’s so excited about it she tells me her results. So I kind of get it secondhand.”

Participants also raised concerns related to AI's misuse and its potential to amplify scams, particularly involving voice simulations. P2 shared a personal experience with a scam attempt using a deep fake voice, stating, “That scam happened to me…it was my granddaughter talking to me, but of course, it's not and so I worry about that.” P7 shared a similar experience about a friend’s near-miss with a voice simulation scam, emphasizing the dangers posed by these technologies when individuals are unaware of their potential for misuse. “ [A resident] here had an occasion where someone called and he was sure that it was his son, even after he hung up, he was still sure that that had been his son, and it was not. Luckily, his wife got a hold of him, and before they did anything”. P9 echoed these concerns, connecting personal experiences with bank fraud to the broader risks AI poses: “We got scammed with our bank account. So people are doing that all the time, and not even using AI. So AI is just adding to that threat, flooding it”.

Participants’ responses varied regarding how the participants currently use GenAI. P12 uses GenAI in his classes he teaches at community college for activities with students to query the bot for a solution to an economic problem. P12 also mentions how it helps him translate medical jargon from a doctor’s diagnosis: “Having it explain my doctor's diagnosis when they give you these cryptic, medical terms, … and it [GenAI] explains it, and then you can ask it to explain how do you treat it?” P4 and P1 indicated searching online and research as their primary use of GenAI while P3 and P6 expressed their intent to leverage it for generating content. P3 stated, “If I have a talk I'm going to give, I'll ask it to take the point to create a PowerPoint.”

To equip themselves with better knowledge and enhance safety in the AI era, P12 emphasized the need for AI literacy education, highlighting its role in empowering knowledge distribution and effective learning: “So I think it's more of a job as those of you who are educators, that are educating the powers to be is that this is an invaluable tool. We need to avail ourselves of it to enhance knowledge distribution”

At the same time some participants expressed concerns about teaching people to become comfortable with AI without sufficiently addressing the risks. For example, P3 highlighted concerns about overfamiliarity with AI leading to reduced vigilance, particularly among vulnerable populations, stating, “You're teaching people to be comfortable with AI, right? And can you sufficiently inoculate against that familiarity they think they know what they're doing to help them be more cautious”. 

\subsubsection{Experience and Feedback on Litti}
The overall sentiment of the Litti was generally positive. One participant found Littie \textit{“really interesting”} (P8); another found the information it provided to their satisfaction, \textit{“Everything that [Litti] gave was, was perfectly normal”} (P5). Participants were particularly impressed when asking the chatbot to conduct creative tasks for them: \textit{“I was fascinated that I could ask it to write a story for my granddaughter…this is really neat.”} (P10)

\textit{“I was very impressed with .. [writing] a birthday greeting. And I was really impressed with results…it was well, well written. It had a nice flow to it. It was very detailed, but yet… it was not a mechanical response, it really sounded like a human had written it.” }(P1)

Furthermore, participants were impressed with the quality of the response given by the chatbot when asked a question about health concerns: \textit{“I am retired eye doctor, and there was a question [about] about dry eyes. [It] gave all of the answers that I would have given to a patient. So I thought it did a very good job.” (}P5)

P8 expressed some ethical concerns about GenAI. Her primary concern was the use of GenAI being used as a substitute for human-writing or people using GenAI to write emails or other documents for them: \textit{“I could be tempted to want to use it, but not always to the advantage of real truth…because, I want to write a little, short essay… and I thought, oh, I could ask AI to write an essay. And that's awful...the temptation to do that was [there].”}

When asked how their perceptions of GenAI changed after using the intervention, P1, P10, and P11 expressed an increase in their understanding and willingness to use it, helping them to see how certain functions can be helpful in their daily tasks. P2, P3, and P8 expressed no change in their perceptions, citing they need \textit{“more time”} to get a better sense of its capabilities while one of them was \textit{“already familiar using them pretty often.”}

When asked about improvements to the chatbot, P1 and P5 specified increasing the font size to 18pt, the standard size at the senior home. P1 also indicated real-life providing examples to older adults who are not very tech-savvy.

\textit{“[For] the folks who are not as tech friendly as we are…just to show them for their awareness, that they're aware that these things are possible, that these things do happen, and we have some live examples just to expose them to it and let them know that be on the lookout for this, this could happen.”}

P6 desired to have the content compacted to reduce the amount of scrolling on the screen. P6 also added a problem with the latency between sending a message to the bot and waiting for a response.

\subsubsection{Trust in Using GenAI}
We asked the participants how their trust in GenAI or willingness to use AI had been affected after Litti experience. Participants expressed a range of perceptions regarding trust in using GenAI, reflecting skepticism and concerns about privacy and security. 

P11 expressed their uncertainty, saying, "I’m still skeptic, but not an agnostic. I’m still waiting to find that magical moment when I say, hey, it makes my life better,... it hasn’t happened yet." 

Similarly, P7 emphasized their reliance on traditional methods, stating they would "consult my doctor first before I would AI"  while also acknowledging that many older adults remain unaware of the dangers associated with AI systems, underscoring the importance of AI literacy to address these gaps, they added, "Yeah, I mean, I do look up lots of things, but I still probably would take it with a grain of salt," conveying a sense of hesitation.

Concerns about security and privacy were common, with P6 expressing uncertainty about data usage, which further reinforced their hesitation to fully embrace GenAI: "You have no guarantee when you enter information on a computer, what happens to it next? So how can you trust when you don’t know where it’s all going?" 

P8 expressed deeper fears, describing AI as "scary" and highlighting "tremendous problems" it could cause. While they reflected on their initial skepticism, noting that when they voiced it, the AI system validated their doubts by responding, "You’re right to have that skepticism". Such interactions seemed to reinforce, rather than alleviate, concerns about the technology.

Some participants also raised ethical and societal concerns, particularly about AI's impact on younger generations, expressing worries about the potential for young people to use the technology irresponsibly. P10 indicated, “I'm still a bit wary…I'm concerned about young people and their use of this technology.” Similarly, P4 accepted the inevitability of AI advancements and trust that AI is part of the future, “All the advertisements and all the things we get are AI driven as today. So as far as trust is concerned, it's a scientific field. It's going to be happening.”

\subsection{Quantitative}
The quantitative analysis examined the impact of Litti on older adults' AI literacy, as well as the specific changes observed in three key sub-scales: AI knowledge, AI detection, and AI ethics. Due to the small sample size, the correlation analysis between AI literacy improvement and factors such as tech self-efficacy, AI familiarity, AI usage, and gender did not yield any statistically significant results. However, the quantitative data showed a positive trend toward improvement in older adults’ overall AI literacy.

\begin{figure*}[htbp!]
 \includegraphics[width=0.5\textwidth]{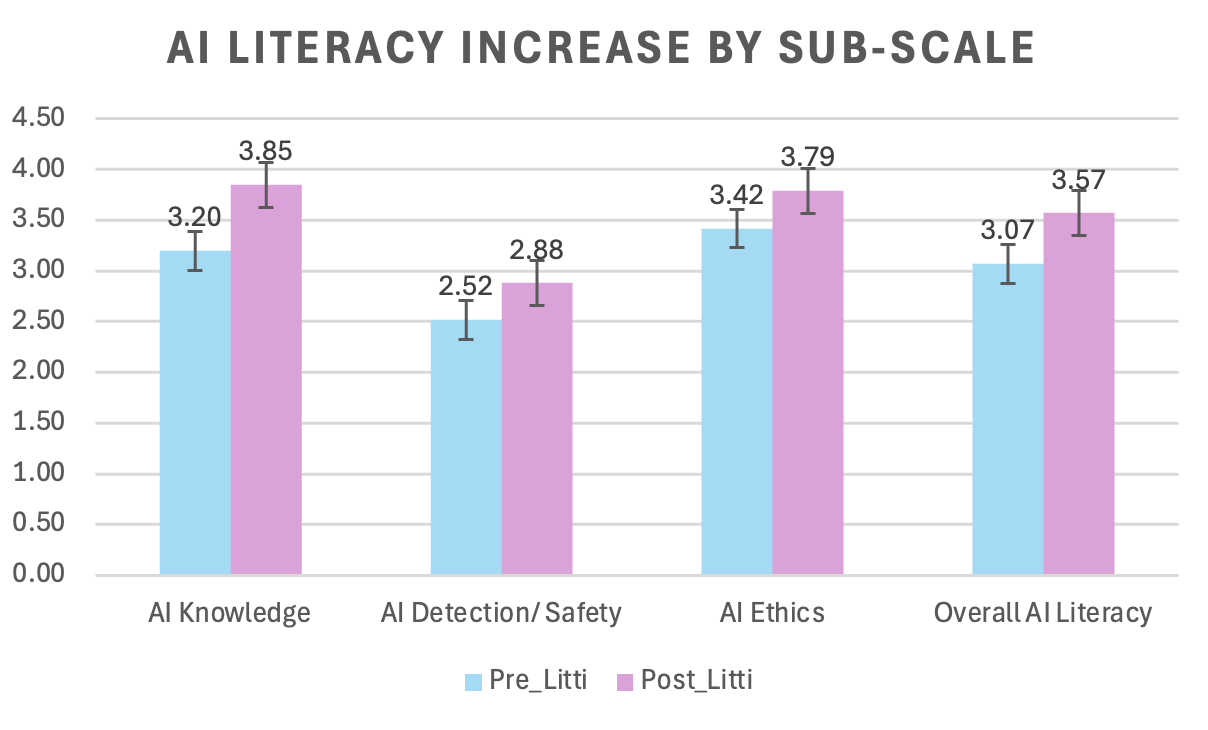}
  \caption{AI literacy increase by individual}
 \label{fig:subscale}
\end{figure*}

The results indicate a meaningful improvement in the participants’ overall digital literacy, with an average increase of 0.55 points (STDEV=0.72) between the pre-test and post-test scores. Sub-scale analyses revealed improvements across the three key areas. Participants demonstrated an increase of 0.72 points (STDEV=0.88) in their AI-related knowledge, suggesting the intervention was effective in enhancing their understanding of AI concepts. The participants exhibited a heightened awareness of the ethical considerations surrounding AI, with a 0.47 point (STDEV=0.91) increase in the AI ethics sub-scale score. Lastly, their ability to detect AI-generated content improved, as evidenced by a 0.37 point (STDEV=1.06) increase in the AI detection sub-scale score. These findings suggest that the Litti equipped the participants with a more comprehensive and nuanced perspective of generative AI and its societal implications (See Figure \ref{fig:subscale}).

Interestingly, the data also revealed that participants with lower pre-literacy scores tended to have larger improvements in their post-literacy scores compared to those with higher pre-literacy scores (See Figure \ref{fig:quant} in Appendix). For example, Participant 10, who had the lowest pre-literacy score of 1.64, showed the greatest improvement of 2.18 points, while Participant 12, who had one of the highest pre-literacy scores of 3.91, exhibited a decline of 0.36 points.

\section{Discussion}
As enAI tools become increasingly integrated into daily life, equipping older adults with the knowledge and skills to navigate this technology is critical \cite{atkinson_breaking_2016, zhang_artificial_2019}. This study explored effective approaches to AI literacy education for older adults, focusing on their awareness, usage, and safety concerns. Our findings highlight the complexities of designing inclusive and impactful AI literacy programs for this demographic, emphasizing the need for tailored, accessible, and trust-building educational strategies.

The study revealed variability in older adults’ familiarity and experience with GenAI. While some had never interacted with AI systems, others had used multiple tools, underscoring the challenge of designing a one-size-fits-all curriculum. Additionally, accessibility issues, such as hearing impairments, highlighted the limitations of voice-based delivery methods. These findings stress the importance of adaptable, multimodal educational approaches that account for varying levels of technological proficiency and physical capabilities. Our findings align with Shandilya and Fan \cite{shandilya_understanding_2022} emphasize the need to design AI technology for older adults by accommodating differences in digital literacy and capacities, such as enabling users to activate or deactivate listening modes in AI-enabled products. Therefore, AI literacy programs must be flexible, offering multiple entry points to accommodate diverse needs. 

The research also uncovered nuanced differences in older adults’ receptiveness to AI literacy education, highlighting the balance on GenAI usage and safety. Participants were relatively more receptive to learning about the usage of GenAI, expressing pleasant surprise at the technology’s ability to assist with daily tasks, such as writing birthday cards or creating customized recipes. However, when it came to AI safety, the intervention seemed insufficient, and participants continued to express concerns related to the misuse of these technologies. This suggests the critical need to strengthen the AI safety-related content within literacy education programs, thoroughly addressing older adults’ valid worries.

A key challenge that emerged from the findings is establishing trust in GenAI. Participants expressed skepticism about the privacy, security, and ethical implications of AI, which poses a significant hurdle to adoption. However, researchers have indicated that raising awareness and providing education about AI can help mitigate these vulnerabilities, given the pervasive role of AI in modern digital applications \cite{zhang_artificial_2019}. As discussed in our interviews, incorporating practical and relatable examples, such as demonstrating tools that detect AI-generated misinformation or exploring tools to identify AI-generated voice fraud, was identified as a crucial strategy to better support non-tech-savvy older adults \cite{xiang_ai_2023}  Addressing these through transparent discussions about AI’s capabilities and limitations must become a central element of AI literacy education.

From our exploration, we highlight several possible design directions for creating similar AI literacy education systems. Compared to traditional digital literacy education methods used for older adults, such as video-based instructions or slide-based workshops, we suggest that chatbots can play a critical role in enhancing AI literacy. This is because chatbots offer interactive, accessible, and adaptive learning experiences, which can be particularly beneficial in educational settings for older adults \cite{pribble_fostering_2024, ng_ai_2022}. They can simplify complex AI concepts, provide real-time feedback, and guide users through structured interactions, reducing cognitive load and improving comprehension \cite{jin_chatting_2024}. 

Furthermore, designing the AI literacy education system as a mobile-friendly chatbot provides flexibility for users with a diverse range of proficiencies with certain devices, such as smartphones and tablets. Providing multimodal input and output will enable users of varying abilities to interact with the system, promoting an accessibility-first approach. The tasks provided to the participants should be designed to provide familiarity with situations more common to older adults, such as health information seeking. Providing scenarios and tasks that are familiar with the target audience will improve the learning process, enabling them to comprehend the knowledge faster. Also, chatbots, when incorporated with ethical AI principles, can encourage critical engagement with AI technologies, raising awareness of bias, fairness, and data privacy, which can make AI literacy education more interactive and effective \cite{kajiwara_ai_2024}.
LLastly, this study has several limitations. Future research should involve larger, more diverse cohorts and allow for extended engagement with AI tools. Additionally, the unique nature of LLM-based interventions necessitates a reevaluation of time allocation strategies, as traditional task-based approaches may not adequately capture the exploratory nature of AI interactions. Third, an experimental design can provide a better reflection on the efficacy of the Litti. To show how a chatbot could be more effective at teaching about itself compared to a website or video, we can set up a between-subject study design with a control group that uses the Litti intervention versus a control group that uses alternative methods, such as a video or website, to provide AI literacy training. Finally, our sample included many highly educated or white participants with relatively higher technology efficacy and prior AI experience. For future, we aim to recruit participants from multiple senior living centers to ensure inclusiveness from diverse racial, educational, and technological backgrounds.

\section{Conclusion}
AI literacy education for older adults must be adaptive, accessible, and trust-focused. Combining hands-on learning with safety education can empower older adults to navigate the GenAI landscape with confidence and curiosity. As AI evolves, so too must our approaches to educating those who stand to benefit most from its transformative potential.


\begin{acks}
We sincerely thank the Westminster residents for their participation in this study. Special thanks to Leon for organizing the participation. We also appreciate the support of UT Austin iSchool for funding this research through the GSLIS Alumni Teaching Fellowship.
\end{acks}

\bibliographystyle{ACM-Reference-Format}
\bibliography{references}

\section*{Appendix}
\appendix

\section{Participant Demographics}
\begin{table}[h]
\centering
\caption{Participant demographics}
\label{tab:demo}
\resizebox{\columnwidth}{!}{
\begin{tabular}{@{}ccccccc@{}}
\toprule
\textbf{PID} & \textbf{Age} & \textbf{Gender} & \textbf{Race} & \textbf{Education} & \textbf{\begin{tabular}[c]{@{}c@{}}AI \\ Familiarity\end{tabular}} & \textbf{\begin{tabular}[c]{@{}c@{}}AI Usage \\ Frequency\end{tabular}} \\ \midrule
P1 & 76 & Man & Hispanic & \begin{tabular}[c]{@{}c@{}}Some college or \\ associate degree\end{tabular} & 3 & 4 \\
P2 & 81 & Man & White & Masters & 3 & 4 \\
P3 & 77 & Woman & White & Masters & 5 & 5 \\
P4 & 90 & Man & Asian & Doctorate & 3 & 5 \\
P5 & 77 & Man & White & Doctorate & 3 & 5 \\
P6 & 85 & Woman & White & Masters & 4 & 1 \\
P7 & 81 & Woman & White & Masters & 3 & 5 \\
P8 & 85 & Woman & White & Masters & 3 & 5 \\
P9 & 80 & Woman & White & Masters & 3 & 5 \\
P10 & 79 & Woman & White & Bachelors & 1 & 1 \\
P11 & 79 & Man & White & Bachelors & 3 & 2 \\
P12 & 78 & Man & White & Doctorate & 4 & 5 \\ \bottomrule
\end{tabular}
}
\end{table}

\section{Focus Group Questions}
\begin{enumerate}
    \item From the generative AI tools you learned about from the study, which tool(s) can you see yourself using in the future? Why?
    \item After learning about genAI, are there situations where you would be concerned or avoid or be afraid using GenAI? If so, why?
    \item What are your thoughts on the development of generative AI to support older adults? In what ways can you see older adults using generative AI?
    \item If any, what did you like or dislike about the learning experience with the chatbot?
    \item If any, what are some ways the chatbot could be improved for a better learning experience?
\end{enumerate}

\section{Additional Figures}
\begin{figure}[hbt]
  \includegraphics[width=0.5\textwidth]{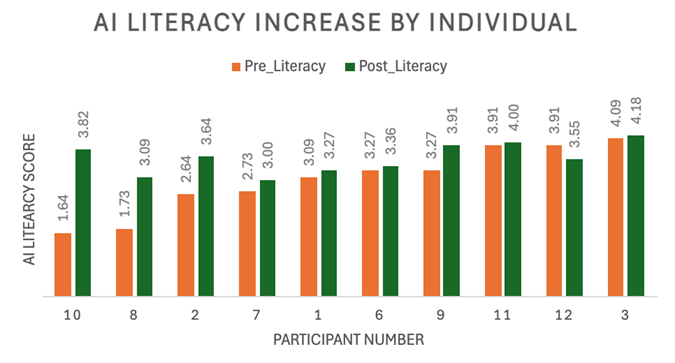}
  \caption{AI literacy increase by individual}
  \label{fig:quant}
\end{figure}

\end{document}